\documentclass[a4paper,twoside,11pt]{article} 
\newlength{\defaultparindent}
\usepackage{latexsym}
\usepackage{amsmath}
\usepackage{amsfonts}
\usepackage{amsthm}
\usepackage{bbold}
\usepackage[applemac]{inputenc} 
\usepackage{multirow}
\usepackage{hyperref}
\hypersetup{colorlinks=true,citecolor=green,linkcolor= blue}
\usepackage{todonotes} 
\usepackage{soul} 

\def\mycc#1{{#1}_{c}} 

\usepackage[arXiv]{optional} 

\opt{AACA}{
\def\cal{\mathcal}
}

\opt{x,AACA}{
\input{mbh_defs_TeX}
}

\opt{arXiv,std,JMP,JOPA}{

\newtheorem{MS_theorem}{Theorem}
\newtheorem{MS_lemma}{Lemma}
\newtheorem{MS_Proposition}{Proposition}
\newtheorem{MS_Corollary}[MS_Proposition]{Corollary}
\def\myconjugate#1{\overline{#1}} 
\def\ie{i.e.\ }
\def\eg{e.g.\ }
\newcommand{\R}{\ensuremath{\mathbb{R}}} 
\newcommand{\C}{\ensuremath{\mathbb{C}}} 
\newcommand{\F}{\ensuremath{\mathbb{F}}} 
\def\my_span#1{\mbox{Span}\left(#1\right)} 

\def\dotinformula{\;\; \mathrm{.}} 

\newcommand{\JJ}{\mathbin{\raisebox{0.25ex}{$\scriptstyle 
                       \rm\vphantom{I}%
                       \_\hskip -0.25em\_%
                       \vrule width 0.6pt$}}}           

\newcommand{\anticomm}[2]{\ensuremath{\left\{ #1, #2 \right\}}} 
\newcommand{\myCl}[3]{\ensuremath{{{\cal C}\ell} {\left( #1, #2 \right)}}}	

}

\opt{JMP}{
}

\begin{document}

\opt{x,std,arXiv,JMP,JOPA}{
\title{{\bf On Spinors of Zero Nullity}
	}

\author{\\
	\bf{Marco Budinich}%
%
%
\\
	University of Trieste and INFN, Trieste, Italy\\
	\texttt{mbh@ts.infn.it}\\
%
%
%
	{\em Advances in Applied Clifford Algebras} (2015)\\
	DOI:10.1007/s00006-015-0547-8\\
	}
\date{ \today }
\maketitle
}

\opt{AACA}{
\title[On Spinors of Zero Nullity]{On Spinors of Zero Nullity}

\author{Marco Budinich}
\address{Dipartimento di Fisica\\
	Università di Trieste \& INFN\\
	Via Valerio 2, I - 34127 Trieste, Italy}
\email{mbh@ts.infn.it}
}

\begin{abstract}
We present a necessary and sufficient condition for a spinor $\omega$ to be of nullity zero, \ie such that for any null vector $v$, $v \omega \ne 0$. This dives deeply in the subtle relations between a spinor $\omega$ and $\mycc{\omega}$, the (complex) conjugate of $\omega$ belonging to the \emph{same} spinor space.
\end{abstract}


\opt{AACA}{
\maketitle
}

\section{Introduction}
\label{Intro}
In 1913 {\'{E}}lie Cartan introduced spinors \cite{Cartan_1913, Cartan_1937} and, more than a century later, this mother lode is far from exhausted. Among spinors simple (pure) spinors are nowadays the least understood. Spinors are deeply intertwined with null (isotropic) vectors and this subject have been visited many times, see \eg \cite{BudinichP_1989, Trautman_Trautman_1994, Charlton_1997, Budinich_2012} and references therein. Here we explore one facet of these relations.

Let the \emph{nullity} $N(\omega)$ of spinor $\omega$ be the dimension of the subspace of null vectors that annihilate $\omega$ \ie those vectors $v$ such that $v \omega = 0$. Simple spinors are the spinors with maximum nullity. Nullity provides a coarse classification of spinors that have been studied in detail: see \cite{Trautman_Trautman_1994, Charlton_1997} and references therein.
In this paper we investigate the properties of a family of spinors complementary to simple spinors: the spinors of zero nullity \ie spinors that are not annihilated by any null vector.

We will investigate these spinors in neutral spaces $\C^{2 m}$ and $\R^{2 m}$ with signature $(m,m)$, a frequent choice in this field \cite{BudinichP_1989, Batista_2014,Trautman_Trautman_1994},
exploiting the Extended Fock Basis (EFB) of Clifford algebra \cite{BudinichM_2009, Budinich_2011_EFB}, recalled in section~\ref{Clifford_algebra_and_EFB}. With this basis \emph{any} element of the algebra can be expressed in terms of simple spinors: from scalars to vectors and multivectors. Section~\ref{Vector_space_V} presents vector and spinor spaces of the algebra and reports some needed results \cite{Budinich_2012}. Section~\ref{spinors_0_nullity} is dedicated to spinors and at the end brings the main result: a necessary and sufficient condition for a spinor to be of zero nullity. With respect to the previous study of this problem \cite{Trautman_Trautman_1994} that tackled Weyl spinors, here the results hold for any spinor. Apart from exceptional cases, a spinor of zero nullity can be seen as the sum of a spinor of positive nullity with its (complex) conjugate. With this result it is easy to build a basis of spinor space made entirely of spinors of zero nullity and also to write down generic spinors with defined nullity.

For the convenience of the reader we tried to make this paper as elementary and self-contained as possible.

\section{Clifford algebra and its 'Extended Fock Basis'}
\label{Clifford_algebra_and_EFB}
We start summarizing the essential properties of the EFB introduced in 2009 \cite{BudinichM_2009, Budinich_2011_EFB}. We consider Clifford algebras \cite{Chevalley_1954} over the field $\F$, with an even number of generators $\gamma_1, \gamma_2, \ldots, \gamma_{2 m}$, a vector space $\F^{2 m} := V$ and a scalar product $g$: these are simple, central, algebras of dimension $2^{2 m}$. As usual
$$
2 g(\gamma_i, \gamma_j) = \gamma_i \gamma_j + \gamma_j \gamma_i := \anticomm{\gamma_i}{\gamma_j} = 2 \delta_{i j} (-1)^{i+1}
$$
and the $\gamma$'s form an orthonormal basis of $V$ with
\begin{equation}
\label{space_signature}
\left\{ \begin{array}{l l l}
\gamma_{2 i - 1}^2 & = & 1 \\
\gamma_{2 i}^2 & = & -1
\end{array} \right.
\qquad i = 1,\ldots,m
\end{equation}
and we concentrate on $\F = \C$ or $\F = \R$ with split signature $V = \R^{m, m}$; given the signature we indicate the Clifford algebra by \myCl{m}{m}{g}.
The Witt, or null, basis of the vector space $V$ is defined, for both $\F = \C$ and $\F = \R$:
\begin{equation}
\label{formula_Witt_basis}
\left\{ \begin{array}{l l l}
p_{i} & = & \frac{1}{2} \left( \gamma_{2i-1} + \gamma_{2i} \right) \\
q_{i} & = & \frac{1}{2} \left( \gamma_{2i-1} - \gamma_{2i} \right)
\end{array} \right.
\Rightarrow
\left\{\begin{array}{l l l}
\gamma_{2i-1} & = & p_{i} + q_{i} \\
\gamma_{2i} & = & p_{i} - q_{i}
\end{array} \right.
\quad i = 1,2, \ldots, m
\end{equation}
that, with $\gamma_{i} \gamma_{j} = - \gamma_{j} \gamma_{i}$, easily gives
\begin{equation}
\label{formula_Witt_basis_properties}
\anticomm{p_{i}}{p_{j}} = \anticomm{q_{i}}{q_{j}} = 0
\qquad
\anticomm{p_{i}}{q_{j}} = \delta_{i j}
\end{equation}
showing that all $p_i, q_i$ are mutually orthogonal, also to themselves, that implies $p_i^2 = q_i^2 = 0$, at the origin of the name ``\emph{null}'' given to these vectors.

Following Chevalley we define spinors as elements of a minimal left ideal we indicate by $S$. Simple spinors are those elements of $S$ that are annihilated by a null subspace of $V$ of maximal dimension.

\bigskip

The EFB of \myCl{m}{m}{g} is given by the $2^{2 m}$ different sequences
\begin{equation*}
\label{EFB_def}
\psi_1 \psi_2 \cdots \psi_m := \Psi \qquad \psi_i \in \{ q_i p_i, p_i q_i, p_i, q_i \} \qquad i = 1,\ldots,m
\end{equation*}
in which each $\psi_i$ can take four different values and we will reserve $\Psi$ for EFB elements. The main characteristics of EFB is that all its elements are \emph{simple} spinors \cite{BudinichM_2009, Budinich_2011_EFB}.

The EFB essentially extends to the entire algebra the Fock basis~\cite{BudinichP_1989} of its spinor spaces and, making explicit the construction $\myCl{m}{m}{g} \cong \overset{m}{\otimes} \myCl{1}{1}{g}$, allows one to prove in \myCl{1}{1}{g} many properties of \myCl{m}{m}{g}.

A classical result we will need in what follows exploits the isomorphism (of vector spaces) $\myCl{m}{m}{g} \cong \Lambda V$ with the Grassmann algebra and leads \cite{Chevalley_1954} to the following useful formula for the Clifford product $v \mu$ of any two elements $v \in V, \mu \in \myCl{m}{m}{g}$
\begin{equation}
\label{Clifford_product}
v \mu := v \JJ \mu + v \wedge \mu
\end{equation}
where $v \JJ \mu$ represents the {\em contraction} of $v$ with $\mu$ (if also $\mu \in V$ then $2 v \JJ \mu = \anticomm{v}{\mu}$) and $v \wedge \mu$ is the {\em exterior} or {\em wedge product}.

\section{Properties of vector $V$ and spinor $S$ spaces}
\label{Vector_space_V}
With the Witt basis (\ref{formula_Witt_basis}) it is easy to see that the null vectors $\{p_{i}\}$ can build vector subspaces made only of null vectors that we call Totally Null Planes (TNP, isotropic planes) of dimension at maximum $m$ \cite{Cartan_1937}. Moreover the vector space $V$ is easily seen to be the direct sum of two of these maximal TNP $P$ and $Q$ respectively:
\begin{displaymath}
V = P \oplus Q \qquad
\left\{ \begin{array}{l l l}
P & := & \my_span{p_1, p_2, \ldots, p_m} \\
Q & := & \my_span{q_1, q_2, \ldots, q_m}
\end{array} \right.
\end{displaymath}
since $P \cap Q = \{0\}$ each vector $v \in V$ may be expressed in the form $v = \sum\limits_{i=1}^{m} \left( \alpha_{i} p_{i} + \beta_{i} q_{i} \right)$ with $\alpha_{i}, \beta_{i} \in \F$.
Using (\ref{formula_Witt_basis_properties}) it is easy to derive the anticommutator of two generic vectors $v$ and $u = \sum\limits_{i=1}^{m} \left( \gamma_{i} p_{i} + \delta_{i} q_{i} \right)$
\begin{equation}
\label{v_u_commutator}
\anticomm{v}{u} = \sum\limits_{i=1}^{m} \alpha_{i} \delta_{i} + \beta_{i} \gamma_{i} \quad \in \F
\quad \Rightarrow \quad
\frac {1}{2} \anticomm{v}{v} = v^2 = \sum\limits_{i=1}^{m} \alpha_{i} \beta_{i} \dotinformula
\end{equation}
We define
$$
V_0 = \{ v \in V : v^2 = 0\} \qquad V_1 = \{ v \in V : v^2 \ne 0\}
$$
clearly $V = V_0 \cup V_1$ and $V_0 \cap V_1 = \emptyset$ but neither $V_0$ nor $V_1$ are subspaces of $V$, which is simple to see. Nevertheless $V_0$ contains subspaces of dimension $m$, \eg $Q$, and, similarly, $V_1 \cup \{ 0 \}$ contains subspaces of dimension $m$, \eg $\my_span{\gamma_1, \ldots, \gamma_{2 k - 1}, \ldots, \gamma_{2 m - 1}}$.
It is well known (and proved explicitly also in \cite{Budinich_2012}) that for any nonzero vector $v$ and spinor $\omega$
\begin{equation}
\label{V0/1_formula}
v \omega = 0 \quad \implies \quad v \in V_0
\end{equation}
and thus, for all $v \in V_1$ and for any $\omega$, $v \omega \ne 0$.

%

\subsection{Conjugation in $V$}
\label{Conjugationi_V}
When $\F = \C$, assuming $\myconjugate{P} = Q$%
\footnote{a sufficient condition is $V = \C \otimes \R^{m,0}$ \cite[p.~35]{BudinichP_1988e} but this condition is not necessary}%
, complex conjugation in vector space $V$ is given by
\begin{equation}
\label{v_vbar}
v = \sum_{i = 1}^m \alpha_i p_i + \beta_i q_i \quad \Rightarrow \quad \myconjugate{v} = \sum_{i = 1}^m \myconjugate{\beta}_i p_i + \myconjugate{\alpha}_i q_i
\end{equation}
that, with (\ref{v_u_commutator}), gives $\myconjugate{v}^2 = \myconjugate{v^2}$.
For $\F = \R$, the field coefficients are real $\myconjugate{\alpha}_i = \alpha_i$, and one can define a similar conjugation that just exchanges basis vectors $p_i$ and $q_i$ (or, identically, exchanges $\alpha_i$ and $\beta_i$)%
\footnote{this conjugation is an $\R$-linear, involutive, automorphism on $\R^{m, m}$ that lifts to the $\C$-linear part of complex conjugation in the ``corresponding'' complex vector space $\C^m$.}%
. In both cases conjugation defines an involutive automorphism on $V$ since $\myconjugate{\myconjugate{v}} = v$.
%
%

For $\F = \R$ we can go further: by (\ref{v_u_commutator}) $\myconjugate{v}^2 = v^2$ and this conjugation is an isometry on $V$ that lifts uniquely to an automorphism on the entire algebra and since our algebra is central simple, all its automorphisms are inner. So there must exist an element $C$ such that $\myconjugate{v} = C v C^{-1}$.

To find its explicit form for our case let $\Delta_\pm = (p_1 \pm q_1) \cdots (p_m \pm q_m)$ and with (\ref{formula_Witt_basis}) it is easy to see that $\Delta_+ = \gamma_1 \cdots \gamma_{2 k - 1} \cdots \gamma_{2 m - 1}$ whereas $\Delta_-$ is the product of the even, spacelike, $\gamma$'s. With (\ref{space_signature}) one easily finds $\Delta_\pm^2 = (-1)^{\frac{m (m \mp 1)}{2}}$ and defining
\begin{equation}
\label{C_def}
C =
\left\{ \begin{array}{l}
\Delta_+\\
\Delta_-
\end{array} \right.
\qquad C^{-1} =
\left\{ \begin{array}{l}
(-1)^{\frac{m (m - 1)}{2}} \Delta_+ \qquad \mbox{for} \; m \; \mbox{odd}\\
(-1)^{\frac{m (m + 1)}{2}} \Delta_- \qquad \mbox{for} \; m \; \mbox{even}
\end{array} \right.
\end{equation}
we can prove that $\myconjugate{v} = C v C^{-1}$: it suffices to write $v$ in the Witt basis and make the simple exercise of proving that $C p_i C^{-1} = q_i$.
One easily verifies
$$
\myconjugate{\myconjugate{v}} = C C v C^{-1} C^{-1} = C C^{-1} v C C^{-1} = v \dotinformula
$$

Returning to the case $\F = \C$, also in this case $C$ is defined and again $C p_i C^{-1} = q_i$ so that, indicating by $v^\star$ the vector $v$ with \emph{complex conjugate field coefficients}, we can write (\ref{v_vbar}) as
\begin{equation*}
\label{v_vbar_generalized}
\myconjugate{v} = C v^\star C^{-1}
\end{equation*}
that holds also for $\F = \R$ since in this case $v^\star = v$ and thus, from now on, we will stick to this form for (complex) conjugation. It is easy to verify that this form generalizes to any element of the algebra $\mu$ giving
$$
\myconjugate{\mu} = C \mu^\star C^{-1}
$$
and that, for both $\F = \C$ and $\R$,
\begin{equation*}
\label{v0_equal_v0bar}
v^2 = 0 \iff \myconjugate{v}^2 = 0
\end{equation*}
and one can prove \cite{Budinich_2012}:
\begin{MS_Proposition}
\label{v_vbar_prop}
Given a nonzero vector $v$ and $\omega \in S$ such that $v \omega = 0$ it follows $\myconjugate{v} \omega \ne 0$, conversely $\myconjugate{v} \omega = 0$ implies $v \omega \ne 0$.
\end{MS_Proposition}

\subsection{Some results for spinor space $S$}
\label{Spinor_spaces}
Given the spinor space $S$ we can build its Fock basis $\Psi_{a}$ where the index $a$ takes $2^m$ values and can be thought expressed in binary form as a string of $m$ ``bits'' taking values $\pm 1$ that represent the $h-$signature of $\Psi_{a}$ \cite{BudinichP_1989, Budinich_2012}. The generic element of $S$ is expressed by the simple spinor expansion:
\begin{equation}
\label{Fock_basis_expansion}
\omega \in S \qquad \omega = \sum_a \xi_{a} \Psi_{a} \dotinformula
\end{equation}


For each nonzero spinor $\omega \in S$ we define its associated TNP as:
$$
M(\omega) := \{v \in V : v \omega = 0\} \qquad \mbox{and} \qquad N(\omega) = \dim_\F M(\omega)
$$
and the spinor is \emph{simple} iff the TNP is of maximal dimension, \ie iff $N(\omega) = m$. A standard result \cite{BudinichP_1988e} says that given $u_1, u_2, \ldots, u_k \in V_0$ they form a TNP of dimension $k$ with $0 < k \le m$ if and only if
\begin{equation}
\label{TNP_properties}
u_1 u_2 \cdots u_k = u_1 \wedge u_2 \wedge \cdots \wedge u_k \ne 0
\end{equation}
that implies also $\anticomm{u_i}{u_j} = 0 \;\; \forall i,j = 1, \ldots, k$ and thus that all vectors in $M(\omega)$ are mutually orthogonal and it is easy to see that $M(\omega)$ is a vector subspace of $V$ contained in $V_0$.

There is also a result \cite{Budinich_2012} complementary to that of proposition~\ref{v_vbar_prop}:
\begin{MS_Proposition}
\label{omega_omegabar_prop}
For any nonzero vector $v$ and $\omega \in S$ such that $v \omega = 0$ it follows $v \myconjugate{\omega} \ne 0$, conversely $v \myconjugate{\omega} = 0$ implies $v \omega \ne 0$.
\end{MS_Proposition}
%
We remark that, given $\omega \in S$, in general $\myconjugate{\omega} = C \omega^\star C^{-1}$ belongs to a different spinor space $S C \ne S$, see \cite{Budinich_2012, Pavsic_2010}. Since $S$ is a minimal left ideal one can define the ``projection'' of $\myconjugate{\omega}$ in the same spinor space of $\omega$ as
\begin{equation}
\label{my_c_def}
\mycc{\omega} := C \omega^\star
\end{equation}
and for any $\omega \in S$ it is simple to get with (\ref{Fock_basis_expansion}) \cite{Budinich_2012}:
\begin{equation}
\label{Fock_basis_expansion_c}
\mycc{\omega} = C \omega^\star = \sum_{a} \myconjugate{\xi}_{a} C \Psi_{a} = \sum_{a} s(a) \myconjugate{\xi}_{a} \Psi_{-a}
\end{equation}
where $s(a) = \pm1$ is a sign, quite tedious to calculate exactly \cite{Budinich_2011_EFB} and $\Psi_{-a}$ is the Fock basis element with $h-$signature opposite to that of $\Psi_{a}$. A significative difference with $\myconjugate{\omega}$ is that while $\myconjugate{\myconjugate{\omega}} = \omega$, $\mycc{(\mycc{\omega})} = C^2 \omega = (-1)^{\frac{m (m-1)}{2}} \omega$. Previous result on $\myconjugate{\omega}$ can be extended \cite{Budinich_2012} to $\mycc{\omega}$:
\begin{MS_Proposition}
\label{omega_omegabar_coro}
For any nonzero $v \in V_0$, for all nonzero $\omega \in S$ such that $v \omega = 0$ it follows $v \mycc{\omega} \ne 0$, conversely $v \mycc{\omega} = 0$ implies $v \omega \ne 0$.
\end{MS_Proposition}
\noindent A useful consequence of this result is:
\begin{equation}
\label{intersections_Ms}
M(\omega) \cap M(\mycc{\omega}) = \{ 0 \} \dotinformula
\end{equation}

\noindent In \cite{Budinich_2012} is proved the
\begin{MS_Proposition}
\label{spinor_def_by_TNP}
Given $k \le m$ nonzero vectors $v_1, v_2, \ldots, v_{k} \in V_0$ forming a TNP of dimension $k$, any spinor that annihilates $v_1, v_2, \ldots, v_{k}$ may be written as
\begin{equation}
\label{General_spinor}
\omega = u_1 u_2 \cdots u_k \Phi
\end{equation}
for an appropriate choice of $\Phi \in S$ whereas the choice of the null vectors $u_i$ is completely free provided they span the same TNP.
\end{MS_Proposition}
\noindent We remark that spinors of the form (\ref{General_spinor}) form vectorial subspaces of $S$ (subsequently called ``pure subspaces'' in \cite{Batista_2014}). We start proving the technical
\begin{MS_lemma}
\label{add_1_to_TNP}
Let $\omega$ be a nonzero spinor with $M(\omega) = \my_span{u_1, u_2, \ldots, u_{N(\omega)}}$, then given a nonzero $v \in V_0$ such that $v \omega \ne 0$ then $N(v \omega) \ge N(\omega)$ equality holding if and only if $\anticomm{v}{u_i} \ne 0$ for at least one $1 \le i \le N(\omega)$
\end{MS_lemma}
\begin{proof}
Spinors are member of a minimal left ideal and thus $v \omega$ is a spinor and $v \in M(v \omega)$. Since $v \omega \ne 0$ then $v \notin M(\omega)$; there are two possibilities: the first is that $v$ is orthogonal to $M(\omega)$, namely $\anticomm{v}{u_i} = 0$ for all $i$, than $M(v \omega)$ contains at least $v$ and all $u_i$ and thus $N(v \omega) > N(\omega)$. If, on the other hand, $\anticomm{v}{u_i} \ne 0$ for at least one $1 \le i \le N(\omega)$, let \eg $\anticomm{v}{u_1} \ne 0$, then $v \omega = (v + u_1) \omega$ but $v + u_1 \in V_1$ and thus is invertible and thus belongs to the Clifford Lipschitz group and it is simple to see that for any of these vectors $M((v + u_1) \omega) = (v + u_1) M(\omega) (v + u_1)^{-1}$ and that $N((v + u_1) \omega) = N(\omega)$.
\end{proof}
In summary $v \omega$ either `adds' $v$ to $M(\omega)$ or `removes' the vector with which $v$ had a nonzero scalar product, neat examples are:
\begin{eqnarray*}
\omega' & = & u_{k+1} \omega = (-1)^k u_1 u_2 \cdots u_k u_{k+1} \Phi \\
\omega' & = & \myconjugate{u}_j \omega = u_1 u_2 \cdots u_{j - 1} \myconjugate{u}_j u_{j + 1} \cdots u_k \Phi'
\end{eqnarray*}
and, in the first case, it is easy to exhibit examples with $N(v \omega) > N(\omega) + 1$.

\section{Spinors of zero nullity}
\label{spinors_0_nullity}
Let's suppose that there are spinors such that
\begin{equation}
\label{ChargeC_eigenvectors}
\mycc{\omega} = \alpha \omega \qquad \alpha \in \F - \{0\}
\end{equation}
by proposition~\ref{omega_omegabar_coro} for these spinors $M(\omega) = \{0\}$, namely $v \omega \ne 0$ for any nonzero $v \in V$. This introduces us to the spinors of zero nullity.

With (\ref{my_c_def}) and (\ref{ChargeC_eigenvectors}), necessarily, $\mycc{(\mycc{\omega})} = \mycc{(\alpha \omega)} = C (\alpha \omega)^\star = \myconjugate{\alpha} \mycc{\omega} = \myconjugate{\alpha} \alpha \omega = |\alpha|^2 \omega = C^2 \omega$, that reduces to $\alpha^2 \omega$ if $\F = \R$ and so, necessarily, in all cases (\ref{ChargeC_eigenvectors}) may hold only if $C^2 = 1$, \ie $m \equiv 0,1 \pmod{4}$%
%
%
\footnote{it is simple to show that $C^2 = -1$ also for real spaces of Lorentzian signature $\R^{2m - 1,1}$}%
. Since $\omega$ and $\mycc{\omega}$ are linearly dependent if and only if (\ref{ChargeC_eigenvectors}) holds, we have proved:
\begin{MS_Proposition}
\label{omega_omega_c}
For $\F = \R$ or $\C$ a nonzero spinor $\omega$ is linearly independent from $\mycc{\omega}$ (\ref{my_c_def}) unless $m \equiv 0,1 \pmod{4}$ and $\mycc{\omega} = \alpha \omega$ with $|\alpha|^2 = 1$.
\end{MS_Proposition}%

The spinors for which $\mycc{\omega} = \alpha \omega$ are the exception, rather than the rule, for spinors of nullity zero. In general spinors $\omega$ and $\mycc{\omega}$ are linearly independent and we will show that, under proper conditions, any of their linear combinations is a spinor of nullity zero; for example let $\omega = q_1 q_2 q_3$ the spinors $\alpha \omega + \beta \mycc{\omega} = \alpha q_1 q_2 q_3 - \beta p_1 q_1 p_2 q_2 p_3 q_3$ has nullity zero for any $\alpha \beta \ne 0$.%

We continue showing that for all spinors $N(\omega) = N(\mycc{\omega})$ (see also \cite{Kopczynski_Trautman_1992}):
\begin{MS_Proposition}
\label{N_N_conjugate}
For any nonzero spinor $\omega \in S$, $N(\omega) = N(\mycc{\omega})$ and if $M(\omega) = \my_span{v_1, v_2, \ldots, v_k}$ then $M(\mycc{\omega}) = \my_span{\myconjugate{v}_1, \myconjugate{v}_2, \ldots, \myconjugate{v}_k}$.
\end{MS_Proposition}
\begin{proof}
Let us suppose first $N(\omega) > 0$, for any $v \in M(\omega)$ one has
$$
0 = v \omega \implies 0 = v^\star \omega^\star = v^\star C^{-1} C \omega^\star \implies 0 = C v^\star C^{-1} C \omega^\star = \myconjugate{v} C \omega^\star = \myconjugate{v} \mycc{\omega}
$$
that implies $N(\mycc{\omega}) \ge N(\omega)$. In turn from $v \in M(\mycc{\omega})$ one has ($C^\star = C$)
$$
0 = v \mycc{\omega} = v C \omega^\star \implies 0 = v^\star C \omega \implies 0 = C^{-1} v^\star C \omega = \myconjugate{v} \omega
$$
that implies $N(\omega) \ge N(\mycc{\omega})$ and thus $N(\omega) = N(\mycc{\omega})$. This argument proves also the part on the composition of TNP's $M(\omega)$ and $M(\mycc{\omega})$.

It remains the case $N(\omega) = 0$: since now $v \omega \ne 0$ for any $v \in V_0$ it follows also $\myconjugate{v} \omega = C^{-1} v^\star C \omega \ne 0$ and this relation can be multiplied by $C$, that, being a product of non null vectors, by (\ref{V0/1_formula}), keeps the result different from zero, thus for any $v \in V_0$ also $v^\star C \omega \ne 0$ and $v C \omega^\star = v \mycc{\omega} \ne 0$ and thus $N(\mycc{\omega}) = 0$.
\end{proof}
\noindent With this proposition applied to (\ref{General_spinor}) we get, for any $0 \le k \le m$
\begin{equation}
\label{General_spinor_and_C}
\omega = u_1 u_2 \cdots u_k \Phi \quad \iff \quad \mycc{\omega} = \myconjugate{u}_1 \myconjugate{u}_2 \cdots \myconjugate{u}_k C \Phi^\star := \myconjugate{u}_1 \myconjugate{u}_2 \cdots \myconjugate{u}_k \mycc{\Phi} \dotinformula
\end{equation}
This result together with (\ref{intersections_Ms}) gives a first characterization of spinors of zero nullity since it is now simple to prove that
$$
N(\omega) = 0 \quad \iff \quad M(\omega) = M(\mycc{\omega})
$$
and clearly (\ref{ChargeC_eigenvectors}) implies $M(\omega) = M(\mycc{\omega})$, not vice versa. To proceed we need some technical results holding for both $\F = \R$ and $\F = \C$:
\begin{MS_lemma}
\label{matched_c_equality_conditions}
For any nonzero spinor $\varphi$ linearly independent from $\mycc{\varphi}$ let
\begin{equation}
\label{generic_tangent_space}
\omega = \alpha \varphi + \beta \mycc{\varphi} \qquad \alpha, \beta \in \F - \{0\}
\end{equation}
then $v \in V_0$ is such that $v \omega = 0$ if and only if
\begin{equation}
\label{matched_spinor_equality}
\alpha v \varphi = - \beta v \mycc{\varphi} \ne 0
\end{equation}
this in turn requires $0 \le N(\varphi) \le 2$. For $N(\varphi) > 0$ necessarily $m > 1$ and, defining $M(\varphi) = \my_span{u_1, u_2, \ldots, u_{N(\varphi)}}$, then $\anticomm{v}{u_i} \ne 0$ and $\anticomm{v}{\myconjugate{u}_j} \ne 0$ for at least one $i$ and one $j$; $i,j = 1, \ldots, N(\varphi)$.
\end{MS_lemma}
\begin{proof}
Given the form of $\omega$, by proposition~\ref{omega_omegabar_coro}, neither $v \varphi$ nor $v \mycc{\varphi}$ can be zero if one wants $v \omega = 0$ that thus can hold only if (\ref{matched_spinor_equality}) holds.

To prove the bounds on $N(\varphi)$ we show that outside these bounds a necessary condition for (\ref{matched_spinor_equality}) does not hold. Let us define spinors $\varphi ' := \alpha v \varphi$ and $\varphi'' := - \beta v \mycc{\varphi}$ with which (\ref{matched_spinor_equality}) reads $\varphi ' = \varphi ''$ that obviously implies
\begin{equation}
\label{necessary_condition}
M(\varphi ') = M(\varphi '') \Rightarrow N(\varphi ') = N(\varphi '')
\end{equation}
moreover $v \in M(\varphi ')$.

If $N(\varphi) = 0$ we have seen that by lemma~\ref{add_1_to_TNP} that $N(\varphi'), N(\varphi'') \ge 1$ and if \eg $M(\varphi ') = M(\varphi '') = \my_span{v}$ then (\ref{necessary_condition}) can be satisfied.

For $N(\varphi) > 0$ with lemma~\ref{add_1_to_TNP} there are four possibilities for $N(\varphi')$ and $N(\varphi'')$ but the two in which $N(\varphi') \ne N(\varphi'')$ are immediately ruled out. There remain either $N(\varphi') = N(\varphi'') \ge N(\varphi) + 1$ or $N(\varphi') = N(\varphi'') = N(\varphi)$. The condition $M(\varphi ') = M(\varphi '')$ with proposition~\ref{N_N_conjugate} rules out the first case since clearly
$\my_span{v, u_1, u_2, \ldots, u_{N(\varphi)}} \ne \my_span{v, \myconjugate{u}_1, \myconjugate{u}_2, \ldots, \myconjugate{u}_{N(\varphi)}}$ for any $N(\varphi) > 0$ so the only remaining possibility is to have $N(\varphi') = N(\varphi'') = N(\varphi)$ that implies, by quoted lemma, $\anticomm{v}{u_i} \ne 0$ and $\anticomm{v}{\myconjugate{u}_j} \ne 0$ for at least one $i, j \in \{ 1, \ldots, N(\varphi) \}$.

We show with an example that if $N(\varphi) = 2$ a solution of (\ref{matched_spinor_equality}) cannot be excluded: let $\varphi = q_1 q_2 \Phi$, $\mycc{\varphi} = \myconjugate{q}_1 \myconjugate{q}_2 \mycc{\Phi} = p_1 p_2 \mycc{\Phi}$ and $v = q_1 + p_2$, clearly $v \in V_0$ and $M(\varphi ') = M(\varphi '') = \my_span{q_1, p_2}$ and (\ref{necessary_condition}) could be satisfied.

Supposing $N(\varphi) > 2$ with lemma~\ref{add_1_to_TNP}, since one can always reduce to the case in which $\anticomm{v}{u_i} \ne 0$ and $\anticomm{v}{\myconjugate{u}_j} \ne 0$ for exactly one $i, j \in \{ 1, \ldots, N(\varphi) \}$,
we would have that in $M(\varphi ')$ necessarily remains at least one $u_i$ that appears as $\myconjugate{u}_i$ in $M(\varphi '')$ and thus (\ref{necessary_condition}) can never be realized with which we proved that necessarily $0 \le N(\varphi) \le 2$.

For $m = 1$ the maximum dimension of a TNP is $1$ but to satisfy $N(\varphi ') = N(\varphi '') = 1$ with lemma~\ref{add_1_to_TNP} one should have $\anticomm{v}{u_1} \ne 0$ and $\anticomm{v}{\myconjugate{u}_1} \ne 0$ that would imply $v^2 \ne 0$ against initial hypothesis of $v \in V_0$ so for $N(\varphi) > 0$ we must necessarily have $m > 1$.
\end{proof}

\begin{MS_Corollary}
\label{zero_nullity_first}
For any spinor $\varphi$ with $N(\varphi) > 2$ then any $\omega \in \my_span{\varphi, \mycc{\varphi}}$ given by (\ref{generic_tangent_space}) with $\alpha \beta \ne 0$ has $N(\omega) = 0$.
\end{MS_Corollary}
\begin{proof}
We start remarking that $N(\varphi) > 2$ implies $m > 2$ and that $\varphi$ is linearly independent from $\mycc{\varphi}$ since, otherwise, $N(\varphi) = 0$. Supposing by absurd that $N(\alpha \varphi + \beta \mycc{\varphi}) > 0$, by lemma~\ref{matched_c_equality_conditions} this would require $0 \le N(\varphi) \le 2$, against hypothesis.
\end{proof}

\subsection{The subspace $S_\omega$}
\label{Plane_S_omega}
We show that every $\omega \in S$ defines uniquely a $2$-dimensional subspace $S_\omega \subseteq S$ that corresponds usually to $\my_span{\omega, \mycc{\omega}}$. Given a nonzero $\omega \in S$ let
\begin{equation}
\label{S_omega_definition}
S_\omega =
\left\{ \begin{array}{lll}
\my_span{\omega, \mycc{\omega}} & \iff & \omega \; \mbox{and} \; \mycc{\omega} \; \mbox{are linearly independent}\\
\my_span{\omega_+, \omega_-} & \iff & \mycc{\omega} = \alpha \omega \; \mbox{(see below)}\\
\end{array} \right.
\end{equation}
and in the first case $S_\omega$ is clearly a two dimensional subspace of $S$.
%
%
%
%
In the second case by proposition~\ref{omega_omega_c} necessarily $m \equiv 0,1 \pmod{4}$ and we have already seen that $N(\omega) = 0$; with a slightly modified (\ref{Fock_basis_expansion}) we can write
\begin{equation}
\label{omega_+-}
\omega = \sum_{a > 0} \xi_{a} \Psi_{a} + \xi_{-a} \Psi_{-a} := \omega_+ + \omega_-
\end{equation}
and by proposition~\ref{symmetric_Fock_expansion} (proved in the Appendix together with some companion propositions) necessarily for any $\xi_{b} \ne 0$ it follows that also $\xi_{-b} \ne 0$ so that $\omega_+$ and $\omega_-$ are both non zero and linearly independent. Moreover by (\ref{Fock_basis_expansion_c}) $\mycc{(\omega_\pm)} = \alpha \omega_\mp$ and in this case we define $S_\omega := \my_span{\omega_+, \omega_-}$.
An example for $m = 1$ is $\omega = q + pq$, clearly $\mycc{\omega} = C \omega^\star = (p + q) \omega = \omega$ and $\omega_+ = q$, $\omega_- = pq$ and in this simple case $S_\omega = S$. A property of $S_\omega$ is
\begin{MS_Proposition}
\label{Conjugate_in_S_omega}
Given a nonzero $\omega$ and its $S_\omega$ (\ref{S_omega_definition}), for all $\varphi \in S_\omega$ it follows that also $\mycc{\varphi} \in S_\omega$.
\end{MS_Proposition}
\begin{proof}
For any $\varphi = \alpha \omega + \beta \mycc{\omega}$, $\alpha, \beta \in \F$, then $\mycc{\varphi} = C^2 \myconjugate{\beta} \omega + \myconjugate{\alpha} \mycc{\omega}$; the other definition of $S_\omega$ is proved similarly.
\end{proof}
\begin{MS_Proposition}
\label{Fock_basis_in_S_omega}
Given a nonzero $\omega$ and its $S_\omega$ (\ref{S_omega_definition}), there always exist $\omega_0, \mycc{\omega_0} \in S_\omega$ such that $N(\omega_0) = N(\mycc{\omega_0}) > 0$.
\end{MS_Proposition}
\begin{proof}
If $N(\omega) > 0$ then $\omega_0 := \omega$ and we are done. Things go similarly in the second case of (\ref{S_omega_definition}) since by proposition~\ref{symmetric_Fock_expansion} $N(\omega_\pm) > 0$. It remains the case $N(\omega) = 0$, in this case with (\ref{omega_+-}) and (\ref{Fock_basis_expansion_c})
\begin{equation*}
\mycc{\omega} = \sum_{a > 0} s(-a) \myconjugate{\xi}_{-a} \Psi_{a} + s(a) \myconjugate{\xi}_{a} \Psi_{-a}
\end{equation*}
and let \eg $\xi_b \ne 0$; by proposition~\ref{symmetric_Fock_expansion} necessarily also $\xi_{-b} \ne 0$ so that choosing $\omega_0 := s(b) \myconjugate{\xi}_{b} \omega - \xi_{-b} \mycc{\omega}$ we get:
$$
\omega_0 = \left[ s(b) \xi_{b} \myconjugate{\xi}_{b} - s(-b) \xi_{-b} \myconjugate{\xi}_{-b} \right] \Psi_{b} +
\sum_{a > 0, \; a \ne b} \cdots
$$
where the field coefficient of $\Psi_{-b}$ is $0$. If $\left[ s(b) \xi_{b} \myconjugate{\xi}_{b} - s(-b) \xi_{-b} \myconjugate{\xi}_{-b} \right] \ne 0$, this violates the necessary condition of proposition~\ref{symmetric_Fock_expansion} for a spinor to be of zero nullity and thus $N(\omega_0) > 0$. If $\left[ s(b) \xi_{b} \myconjugate{\xi}_{b} - s(-b) \xi_{-b} \myconjugate{\xi}_{-b} \right] = 0$ one can repeat the procedure starting from the newly defined $\omega_0$ and $\mycc{\omega_0}$ that must be nonzero because, otherwise, the initial spinors $\omega$ and $\mycc{\omega}$ would be linearly dependent. This linear independence guarantees also that this iterative procedure must terminate with the vanishing of just one term because, otherwise, again, the initial spinors would be linearly dependent.
\end{proof}

Beyond the formal proof one can get an intuition of this result from an interesting property of $S_\omega$. The spinor $\omega$ is nonzero, so let us suppose that in its Fock basis expansion (\ref{Fock_basis_expansion}) appears the term $\xi_a \Psi_a$. In the spinor space $S' \ne S$ of $g-$signature $-a$, $\Psi_a$ is a primitive idempotent \cite{Budinich_2011_EFB}. It is not difficult to see that in this spinor space the spinors $\omega, \mycc{\omega}, \mycc{\omega} C^{-1} (= \myconjugate{\omega})$ and $\omega C^{-1}$ (the last two are in $S' C^{-1}$) form a sub algebra of $\myCl{m}{m}{g}$ isomorphic to $\myCl{1}{1}{g}$. So it is always possible to ``rotate'' the minimal left ideal formed by $\omega, \mycc{\omega}$, combining them linearly, to build a Fock basis of $\myCl{1}{1}{g}$ made of two spinors of positive nullity.

We will call the spinors $(\omega_0, \mycc{\omega_0})$ the Fock basis of $S_\omega$; a useful consequence is:
\begin{MS_Corollary}
\label{canonical_expansion}
Given a nonzero $\omega$ and its $S_\omega$ (\ref{S_omega_definition}), any $\varphi \in S_\omega$ can be expressed as $\varphi = \alpha \omega_0 + \beta \mycc{\omega_0}$, $\alpha, \beta \in \F$, with $N(\omega_0) = N(\mycc{\omega_0}) > 0$.
\end{MS_Corollary}

For the next proposition, bringing the main result, we need a different form for the generic spinor $\omega \in S$ that exploits the properties of the Fock basis expansion (\ref{Fock_basis_expansion}). If $m \ge 2$ one can collect all terms with identical first two components of (\ref{Fock_basis_expansion}) and any $\omega \in S$ may be written as
\begin{equation}
\label{m-m-2_form}
\omega = q_1 q_2 \Phi_{q q} + q_1 p_2 q_2 \Phi_{q p} + p_1 q_1 q_2 \Phi_{p q} + p_1 q_1 p_2 q_2 \Phi_{p p}
\end{equation}
where the spinors $\Phi_{x y}$ belong to a spinor space $S'$ of dimension $2^{m - 2}$ and contain all the field coefficients $\xi_a$ of (\ref{Fock_basis_expansion}). We remark the subtle difference with (\ref{General_spinor_and_C}): whereas there $\Phi \in S$ and the relation works since $S$ is a minimal left ideal, here $\Phi_{x y} \in S'$ and we are exploiting the properties of Fock basis expansion (\ref{Fock_basis_expansion}). The difference emerges when we calculate $\mycc{\omega}$: writing from (\ref{C_def}) $C = \left(p_1 + (-1)^{m - 1} q_1\right) \cdots \left(p_m + (-1)^{m - 1} q_m\right)$, we find from (\ref{my_c_def})
\begin{eqnarray*}
\mycc{\omega} & = & C q_1^\star q_2^\star \Phi_{q q}^\star + C q_1^\star p_2^\star q_2^\star \Phi_{q p}^\star + C p_1^\star q_1^\star q_2^\star \Phi_{p q}^\star + C p_1^\star q_1^\star p_2^\star q_2^\star \Phi_{p p}^\star
\end{eqnarray*}
and we observe that $q_i^\star = q_i$ because they all have field coefficients $1$ (all field coefficients that are not $1$ are actually buried in $\Phi_{x y}$) and defining $C' := \left(p_3 + (-1)^{m - 1} q_3\right) \cdots \left(p_m + (-1)^{m - 1} q_m\right)$ the conjugation operator of the spinor space $S'$ we find (obviously $(-1)^{m - 3} = (-1)^{m - 1}$)
\begin{eqnarray}
\label{m-m-2_form_c}
\mycc{\omega} & = & C q_1 q_2 \Phi_{q q}^\star + C q_1 p_2 q_2 \Phi_{q p}^\star + C p_1 q_1 q_2 \Phi_{p q}^\star + C p_1 q_1 p_2 q_2 \Phi_{p p}^\star = \nonumber \\
& = & - \left(p_1 + (-1)^{m - 1} q_1\right) q_1 \left(p_2 + (-1)^{m - 1} q_2\right) q_2 C' \Phi_{q q}^\star + \nonumber \\
& & + (-1)^{m - 1} \left(p_1 + (-1)^{m - 1} q_1\right) q_1 \left(p_2 + (-1)^{m - 1} q_2\right) p_2 q_2 C' \Phi_{q p}^\star + \nonumber \\
& & + (-1)^{m - 2} \left(p_1 + (-1)^{m - 1} q_1\right) p_1 q_1 \left(p_2 + (-1)^{m - 1} q_2\right) q_2 C' \Phi_{p q}^\star + \nonumber \\
& & + \left(p_1 + (-1)^{m - 1} q_1\right) p_1 q_1 \left(p_2 + (-1)^{m - 1} q_2\right) p_2 q_2 C' \Phi_{p p}^\star = \nonumber \\
& = & q_1 q_2 \mycc{\Phi_{p p}} - q_1 p_2 q_2 \mycc{\Phi_{p q}} + p_1 q_1 q_2 \mycc{\Phi_{q p}} - p_1 q_1 p_2 q_2 \mycc{\Phi_{q q}} \dotinformula
\end{eqnarray}

\subsection{The case of $N(\omega_0) \le 2$}
Given $\omega$ and its $S_\omega$ we give now sufficient conditions for having spinors of nullity zero in the case that the Fock basis of $S_\omega$ has $N(\omega_0) \le 2$:
\begin{MS_Proposition}
\label{S_omega_nullity_0}
Given a nonzero $\omega$ and its $S_\omega$ (\ref{S_omega_definition}) with its Fock basis $(\omega_0, \mycc{\omega_0})$ and $m > 2$, then for $m \equiv 0, 1 \pmod{4}$ or $N(\omega_0) > 2$, for any $\varphi = \alpha \omega_0 + \beta \mycc{\omega_0}$, $\alpha, \beta \in \F$ and $\alpha \beta \ne 0$, then $N(\varphi) = 0$.

\noindent For $m \equiv 2, 3 \pmod{4}$ and $N(\omega_0) \le 2$ additional conditions are needed on the $\Phi_{x y} \in S'$ of the expansion (\ref{m-m-2_form}) of $\omega_0$, namely:
\begin{itemize}
\item if $N(\omega_0) = 2$, let $\omega_0 = q_1 q_2 \Phi_{q q}$ (not a limitation, see proof) than to have $N(\varphi) = 0$ $\Phi_{q q}$ must be linearly independent from $\mycc{\Phi_{q q}}$;
\item if $N(\omega_0) = 1$, let $\omega_0 = q_1 q_2 \Phi_{q q} + q_1 p_2 q_2 \Phi_{q p}$ (again, not a limitation) than to have $N(\varphi) = 0$ at least one of the three following conditions must be satisfied: $\Phi_{q q}$ is linearly independent from $\mycc{\Phi_{q q}}$, $\Phi_{q p}$ is linearly independent from $\mycc{\Phi_{q p}}$ and $|\alpha|^2 \ne |\beta|^2$.
\end{itemize}
\end{MS_Proposition}
\begin{proof}
If $N(\omega_0) = N(\mycc{\omega_0}) > 2$ we already know, by corollary~\ref{zero_nullity_first}, that any $\varphi = \alpha \omega_0 + \beta \mycc{\omega_0}$ with $\alpha \beta \ne 0$ has nullity zero; we prove now that this also holds for $N(\omega_0) = N(\mycc{\omega_0}) = 1, 2$ with additional conditions if $m \equiv 2, 3 \pmod{4}$. First of all we note that since $N(\omega_0) > 0$, by lemma~\ref{matched_c_equality_conditions}, $m > 1$.

Let us consider first $N(\omega_0) = N(\mycc{\omega_0}) = 2$ and let $M(\omega_0) = \my_span{u_1, u_2}$ for some $u_1, u_2 \in V_0$.
Without loss of generality we can assume $M(\omega_0) = \my_span{q_1, q_2}$, since it is always possible to make a proper rotation in the Witt basis (\ref{formula_Witt_basis}) to get this; so we can write, with (\ref{m-m-2_form}) and (\ref{m-m-2_form_c})
\begin{equation*}
\label{Fock_basis_S_omega}
\omega_0 = q_1 q_2 \Phi_{q q} \qquad \mycc{\omega_0} = - p_1 q_1 p_2 q_2 \mycc{\Phi_{q q}} \dotinformula
\end{equation*}
We proceed by contradiction supposing that there exists $v \in V_0$ such that $v (\alpha \omega_0 + \beta \mycc{\omega_0}) = 0$. By necessary conditions of lemma~\ref{add_1_to_TNP} we must have $\anticomm{v}{q_i} \ne 0$ and $\anticomm{v}{p_j} \ne 0$ with $1 \le i,j \le 2$ and there are two possibilities: the first is $i = j$; in this case we may always write, in full generality
$$
v = q_i + \xi p_i + v' \qquad 1 \le i,j \le 2, \;\; \xi \in \F
$$
with $\anticomm{v'}{q_i} = \anticomm{v'}{p_i} = 0$ and $v'^2 = - (q_i + \xi p_i)^2 = - \xi$ and, since we can always obtain that $v$ has nonzero scalar product with just one $q_i$ and one $p_i$, we can conclude that also for the other coordinate $\anticomm{v'}{q_j} = \anticomm{v'}{p_j} = 0$. It is easy to see that in this case, supposing \eg $i = 1$, $M(v \omega_0) = \my_span{v, q_2}$ while $M(v \mycc{\omega_0}) = \my_span{v, p_2}$ that violates necessary conditions (\ref{necessary_condition}) and so in this case $v (\alpha \omega_0 + \beta \mycc{\omega_0}) \ne 0$. The second possibility is that $i \ne j$ and let \eg $\anticomm{v}{q_2} \ne 0$ and $\anticomm{v}{p_1} \ne 0$; it follows that we may write
$$
v = q_1 + \xi p_2 + v' \qquad \xi \in \F
$$
and again $\anticomm{v'}{q_1} = \anticomm{v'}{p_1} = \anticomm{v'}{q_2} = \anticomm{v'}{p_2} = 0$ and in this case $v'^2 = 0$; we get now
$$
(q_1 + \xi p_2 + v') (\alpha \omega_0 + \beta \mycc{\omega_0}) = \alpha \xi p_2 \omega_0 + \alpha v' \omega_0 + \beta q_1 \mycc{\omega_0} + \beta v' \mycc{\omega_0}
$$
and since $v' \omega_0 \ne 0$ and $v' \mycc{\omega_0} \ne 0$ by the hypothesis $N(\omega_0) = 2$ we must conclude that, to satisfy the relation, one must necessarily have $v' = 0$ because there are no other ways that the terms $\alpha v' \omega_0$ and $\beta v' \mycc{\omega_0}$ can cancel out. So the relation reduces to $\alpha \xi p_2 \omega_0 + \beta q_1 \mycc{\omega_0} = 0$ where both terms are again nonzero and it is easy to see that
$$
\alpha \xi p_2 \omega_0 + \beta q_1 \mycc{\omega_0} = - q_1 p_2 q_2 (\alpha \xi \Phi_{q q} + \beta \mycc{\Phi_{q q}}) = 0
$$
and we observe that $q_1 p_2 q_2 \ne 0$ and the term in parenthesis is a spinor in $S'$ that cannot be brought to zero by any of the null vectors that precedes it. So this expression can be zero only if, in $S'$ spinor space,
\begin{equation}
\label{S'equation}
\mycc{\Phi_{q q}} = - \frac{\alpha \xi}{\beta} \Phi_{q q} \dotinformula
\end{equation}
We remark that if $m = 2$ this expression involves only field coefficients and can thus \emph{always} be solved to zero; this shows that there are no spinors of zero nullity in this case, an anticipation of a more general result proved later.

Since, by hypothesis, $m > 2$ then, by proposition~\ref{omega_omega_c} (\ref{S'equation}) can have solution only for $m - 2 \equiv 0, 1 \pmod{4}$ \ie $m \equiv 2, 3 \pmod{4}$ with the necessary condition $|\alpha|^2 |\xi|^2 = |\beta|^2$ that shows that for any $\alpha, \beta$ the vector $v = q_1 + \xi p_2$, with a value of $\xi$ satisfying $|\xi|^2 = |\beta|^2 / |\alpha|^2$, annihilates $\alpha \omega_0 + \beta \mycc{\omega_0}$. So to have $N(\varphi) = 0$ we must add the additional condition that $\Phi_{q q}$ is linearly independent from $\mycc{\Phi_{q q}}$ (that is automatically satisfied if \eg $N(\Phi_{q q}) > 0$ that happens, for example, when $N(\omega_0) > 2$).

\bigskip

We go now to the case $N(\omega_0) = N(\mycc{\omega_0}) = 1$ and, by the same hypothesis of previous case, we can assume $M(\omega_0) = q_1$ and we can write, with (\ref{m-m-2_form}) and (\ref{m-m-2_form_c}) and in full generality
\begin{eqnarray*}
\omega_0 & = & q_1 q_2 \Phi_{q q} + q_1 p_2 q_2 \Phi_{q p} \\
\mycc{\omega_0} & = & p_1 q_1 q_2 \mycc{\Phi_{q p}} - p_1 q_1 p_2 q_2 \mycc{\Phi_{q q}} \dotinformula
\end{eqnarray*}
We proceed again by contradiction supposing that there exists $v \in V_0$ such that $v (\alpha \omega_0 + \beta \mycc{\omega_0}) = 0$. By necessary conditions of lemma~\ref{add_1_to_TNP} we must have $\anticomm{v}{q_1} \ne 0$ and $\anticomm{v}{p_1} \ne 0$ so that we may always write in full generality
$$
v = q_1 + \xi p_1 + v' \qquad \xi \in \F
$$
with $\anticomm{v'}{q_1} = \anticomm{v'}{p_1} = 0$ and since $v$ is null we must have $v'^2 = - (q_1 + \xi p_1)^2 = - \xi$ so that
$$
v \varphi = (q_1 + \xi p_1 + v') (\alpha \omega_0 + \beta \mycc{\omega_0}) = \alpha (\xi p_1 + v') \omega_0 + \beta (q_1 + v') \mycc{\omega_0}
$$
and we observe that $(\xi p_1 + v')^2 = (q_1 + v')^2 = v'^2 = -\xi $ and thus, by (\ref{V0/1_formula}), both terms in the equality are nonzero so that, to satisfy $v \varphi = 0$, one must have
$$
\mycc{\omega_0} = \frac{\alpha}{\beta \xi} (q_1 + v') (\xi p_1 + v') \omega_0 = \cdots = \frac{\alpha}{\beta} v' p_1 \omega_0 \dotinformula
$$
We observe now that the only request made on $v'$ is that it must be orthogonal to the subspace $\my_span{q_1, p_1}$ so that it is always possible to make a proper rotation in $V$ basis to obtain, without loss of generality, that
$$
v' = q_2 - \xi p_2
$$
with which the necessary condition becomes:
\begin{eqnarray*}
p_1 q_1 q_2 \mycc{\Phi_{q p}} - p_1 q_1 p_2 q_2 \mycc{\Phi_{q q}} & = & \frac{\alpha}{\beta} (q_2 - \xi p_2) p_1 (q_1 q_2 \Phi_{q q} + q_1 p_2 q_2 \Phi_{q p}) = \\
& = & \frac{\alpha}{\beta} ( p_1 q_1 q_2 \Phi_{q p} - \xi p_1 q_1 p_2 q_2 \Phi_{q q})
\end{eqnarray*}
that, to be satisfied, needs that two equations are separately satisfied
\begin{eqnarray*}
p_1 q_1 q_2 (\mycc{\Phi_{q p}} - \frac{\alpha}{\beta} \Phi_{q p}) = 0\\
p_1 q_1 p_2 q_2 (\mycc{\Phi_{q q}} - \frac{\alpha}{\beta} \xi \Phi_{q q}) = 0
\end{eqnarray*}
and again for $m > 2$ these equations can be satisfied only for $m \equiv 2, 3 \pmod{4}$ and in this case, if $|\alpha|^2 = |\beta|^2$, it is always possible to find $v$ such that $v \varphi = 0$. Consequently to get $N(\varphi) = 0$ it is sufficient that either $\Phi_{q q}$ is linearly independent from $\mycc{\Phi_{q q}}$ or $\Phi_{q p}$ from $\mycc{\Phi_{q p}}$ or that $|\alpha|^2 \ne |\beta|^2$.
\end{proof}

\subsection{The main result}
We resume all previous results in the following characterization of spinors of zero nullity:
\begin{MS_theorem}
\label{zero_nullity_theorem}
In $\myCl{m}{m}{g}$ with $m \ne 2$ a nonzero spinor $\omega \in S$ has $N(\omega) = 0$ if and only if it can be written in the Fock basis $(\omega_0, \mycc{\omega_0})$ of its $S_\omega$ (\ref{S_omega_definition}) as
\begin{equation*}
\omega = \alpha \omega_0 + \beta \mycc{\omega_0} \quad \alpha, \beta \in \F - \{0\} \dotinformula
\end{equation*}

\noindent For $m \equiv 2, 3 \pmod{4}$ and $N(\omega_0) \le 2$ additional conditions are needed on the $\Phi_{x y} \in S'$ of the expansion (\ref{m-m-2_form}) of $\omega_0$, namely:
\begin{itemize}
\item if $N(\omega_0) = 2$, let $\omega_0 = q_1 q_2 \Phi_{q q}$ than to have $N(\omega) = 0$ $\Phi_{q q}$ must be linearly independent from $\mycc{\Phi_{q q}}$;
\item if $N(\omega_0) = 1$, let $\omega_0 = q_1 q_2 \Phi_{q q} + q_1 p_2 q_2 \Phi_{q p}$ than to have $N(\omega) = 0$ at least one of the three following conditions must be satisfied: $\Phi_{q q}$ is linearly independent from $\mycc{\Phi_{q q}}$, $\Phi_{q p}$ is linearly independent from $\mycc{\Phi_{q p}}$ and $|\alpha|^2 \ne |\beta|^2$.
\end{itemize}

The case $m = 2$ is exceptional since there are no spinors of zero nullity.
\end{MS_theorem}
\begin{proof}
Proposition~\ref{S_omega_nullity_0} proves the forward part of the theorem for $m > 2$. We now suppose $N(\omega) = 0$: we can define $S_\omega$ with its Fock basis $(\omega_0, \mycc{\omega_0})$ and obviously $\omega = \alpha \omega_0 + \beta \mycc{\omega_0}$ with $\alpha \beta \ne 0$ because otherwise one would contradict the hypothesis $N(\omega) = 0$. In the particular case $m \equiv 2, 3 \pmod{4}$ and $N(\omega_0) \le 2$ then at least one of the $\Phi_{x y} \in S'$ of its expression (\ref{m-m-2_form}) is linearly independent from its conjugate $\mycc{\Phi_{x y}}$ because otherwise, as pointed out in the proof of proposition~\ref{S_omega_nullity_0}, there always exists a null vector that annihilates $\alpha \omega_0 + \beta \mycc{\omega_0}$ that would contradict our initial hypothesis.

The case $m = 1$ cannot be derived by proposition~\ref{S_omega_nullity_0} but it can be proved directly solving $v \omega = 0$ for the generic null vector and the generic spinor
$$
v \omega = (\alpha p + \beta q) (\xi_1 q + \xi_2 p q) = \beta \xi_2 q + \alpha \xi_1 p q = 0 \qquad \alpha \beta = 0
$$
that can be solved only if $\xi_1 \xi_2 = 0$.

In the case $m = 2$ we already saw in the proof of proposition~\ref{S_omega_nullity_0} that there are no spinors of $0$ nullity but also in this case we can give a direct proof; we can write the generic spinor (\ref{Fock_basis_expansion}) as
$$
\omega = \xi_1 q_1 q_2 + \xi_2 q_1 p_2 q_2 + \xi_3 p_1 q_1 q_2 + \xi_4 p_1 q_1 p_2 q_2
$$
and it is a simple exercise to check that the vector%
\footnote{this is the solution when, $\forall i, \; \xi_i \ne 0$, in other cases it takes slightly different forms.}%
$$
v = \xi_3 \xi_4 p_1 - \xi_1 \xi_2 q_1 - \xi_2 \xi_4 p_2 - \xi_1 \xi_3 q_2
$$
is null and such that $v \omega = 0$.
\end{proof}

\section{Conclusions}
The first offspring of this result is that one can build a basis of spinor space(s) $S$ made entirely of spinors of zero nullity since, from (\ref{Fock_basis_expansion}) one can write
\begin{eqnarray*}
\omega & = & \sum_{a > 0} \xi_{a} \Psi_{a} + \xi_{-a} \Psi_{-a} = \\
& = & \sum_{a > 0} \frac{\xi_{a} + \xi_{-a}}{2} (\Psi_{a} + \Psi_{-a}) + \frac{\xi_{a} - \xi_{-a}}{2} (\Psi_{a} - \Psi_{-a})
\end{eqnarray*}
and for $m \ne 2$ the basis $\{ \Psi_{a} + \Psi_{-a}, \Psi_{a} - \Psi_{-a}: a > 0\}$ is made entirely of spinors of zero nullity, each element being the sum of two simple spinors. Moreover any nonzero $\omega$ with $N(\omega_0) = N(\mycc{\omega_0}) > 0$ can be written, not uniquely, as a linear combination of two zero nullity spinors taken from its $S_\omega$.

Another possibly interesting application of this result is that it allows to write down explicitly spinors with a \emph{defined} nullity: generalizing (\ref{m-m-2_form}) one can write
$$
\omega = q_1 q_2 \cdots q_k \Phi_{q}
$$
where the spinor $\Phi_{q}$ belong to a spinor space $S'$ of dimension $2^{m - k}$ and while $\Phi_{q}$ 'spans' its spinor space $S'$, the nullity of $\omega$ 'spans' the interval $[k, m]$ and so the only thing that one can say about the nullity of $\omega$ is that $N(\omega) \ge k$. With theorem~\ref{zero_nullity_theorem} one can impose that $N(\Phi_{q}) = 0$ in $S'$ and this guarantees that $N(\omega) = k$; this can be useful in the classification of spinors based on nullity \cite{Trautman_Trautman_1994}.

These results show also the complementary roles of $\omega$ and $\mycc{\omega}$ and that their span contains all spinors of zero nullity but for two ``directions'', those of the Fock basis of $S_\omega$ (apart from pathological cases). This resembles closely the spinor space $S$ of $\myCl{1}{1}{g}$ that has two directions, $q$ and $pq$, of nullity 1 (by the way in this case these are also the simple spinors of $S$) while all other spinors of $S$ are of zero nullity.


\subsection*{Appendix}
Here there are some technical results used in section~\ref{Plane_S_omega}.%
\begin{MS_Proposition}
\label{zero_nullity_expansion}
For all $\omega \in S$ with $N(\omega) = 0$ and any $v \in V_0$ it is always possible to write $\omega$ as:
\begin{equation}
\label{zero_nullity_expansion_formula}
\omega = v \Phi_v + \myconjugate{v} \Phi_{\myconjugate{v}}
\end{equation}
where $\Phi_v, \Phi_{\myconjugate{v}} \in S$ and are both nonzero.
\end{MS_Proposition}
\begin{proof}
Given any couple of null vectors $v, \myconjugate{v}$ it is always possible to make a proper rotation in the Witt basis (\ref{formula_Witt_basis}) to get \eg $v \propto q_1, \myconjugate{v} \propto p_1$ and then the $\omega$ expansion (\ref{zero_nullity_expansion_formula}) is just the Fock basis expansion (\ref{Fock_basis_expansion}) split in the two parts with same first component. Since $N(\omega) = 0$ clearly $v \omega \ne 0$ and $\myconjugate{v} \omega \ne 0$ and if either of $\Phi_v, \Phi_{\myconjugate{v}}$ would vanish this would contradict $N(\omega) = 0$.
\end{proof}

\begin{MS_Proposition}
Given a maximal TNP $V_a \subset V_0$ and its corresponding simple spinor $\Psi_a$, \ie such that $M(\Psi_a) = V_a$, a spinor $\omega \in S$ is such that, for every $v \in V_a$, we can find $\omega' \in S$ giving
$$
\omega = v \omega'
$$
if and only if $\omega = \xi \Psi_a$, with $\xi \in \F$.
\end{MS_Proposition}
\begin{proof}
Since from any maximal TNP we can build a Witt basis of $V$ naming its null vectors $q_i$, without loss of generality we suppose $V_a = Q = \my_span{q_1, \ldots, q_m}$ and $\Psi_a = q_1 q_2 \cdots q_m$.

Supposing first $\omega = \xi \Psi_a$, for any $v = \sum_{i = 1}^m \alpha_i q_i \in Q$ we have
$$
\omega = \xi \Psi_a = \frac{\xi}{m} \left( \sum_{i =1}^m \alpha_i q_i \right) \sum_{i = 1}^m \frac{s(i)}{\alpha_i} \Psi_{a(i)} \qquad \Psi_{a(i)} = q_1 q_2 \cdots p_i q_i \cdots q_m
$$
where $s(i) = \pm 1$ and such that $s(i) q_i \Psi_{a(i)} = \Psi_{a}$ and we have supposed, for simplicity, that all $\alpha_i \ne 0$ (the formula can be easily adapted to other cases).

Conversely let us suppose that $\omega = v \omega'$ for any $v \in Q$, it follows that for any $v \in Q$ one has $v \omega = 0$ that means that $\omega$ is a simple spinor and, by proposition~6 of \cite{Budinich_2012}, $\omega = \xi \Psi_a$ for some $\xi$.
\end{proof}
\noindent This result can be generalized from the case of a simple spinor $\Psi_a$ to the case of a spinor that \emph{contains} $\Psi_a$ in its Fock basis expansion (\ref{Fock_basis_expansion})
\begin{MS_Corollary}
Given a maximal TNP $V_a \subset V_0$ and its corresponding simple spinor $\Psi_a$, a spinor $\omega \in S$ is such that, for every $v \in V_a$, we can find $\omega', \omega'' \in S$, with $\omega' \ne 0$, giving
$$
\omega = v \omega' + \omega''
$$
if and only if $\omega = \xi \Psi_a + \omega'''$ for some $\omega''' \in S$, with $\xi \in \F$.
\end{MS_Corollary}
\begin{proof}
Supposing $\omega = \xi \Psi_a + \omega'''$ previous proposition gives the result. Conversely let $\omega = v \omega' + \omega''$ for any $v \in Q$ (as before we take $V_a = Q$); in this case we proceed by induction on the dimension $m$: for $m = 1$ the most general spinor takes the form $\omega = \xi_1 q + \xi_2 p q$ and the result is obvious. Let us now suppose the proposition to be true for $m - 1$ and let us move to $m$: with self explanatory notation in this case the most general spinor has the form $\omega = q_1 \Phi_q + p_1 q_1 \Phi_p$ and any null vector of $Q$ may be written as $v = \alpha q_1 + \beta q'$ where $q'$ is a null vector of the $m - 1$ dimensional maximal TNP $Q'$. By the induction hypothesis for any null vector $q' \in V'$ we can write $\Phi_q = q' \Phi_q' + \Phi_q''$ and the first term contains the simple spinor $\xi q_2 \cdots q_m$. It follows that our spinor of the case $m$ can be written
$$
\omega = q_1 \Phi_q + p_1 q_1 \Phi_p = q_1 (q' \Phi_q' + \Phi_q'') + p_1 q_1 \Phi_p = (\alpha q_1 + \beta q') \frac{1}{\alpha} q' \Phi_q' + q_1 \Phi_q'' + p_1 q_1 \Phi_p
$$
and thus in the term $q_1 \Phi_q$ appears the simple spinor $\xi q_1 q_2 \cdots q_m$.
\end{proof}

\begin{MS_Proposition}
\label{symmetric_Fock_expansion}
For all $\omega \in S$ with $N(\omega) = 0$, given any $\xi_a \ne 0$ in its expansion (\ref{Fock_basis_expansion}) necessarily also $\xi_{-a} \ne 0$
\end{MS_Proposition}
\begin{proof}
Given any $\xi_a \ne 0$ we write $\omega = \xi_a \Psi_a + \omega'$ and since, by proposition~\ref{zero_nullity_expansion}, for any null vector $v \in M(\Psi_a)$ we can write $\omega$ also as in (\ref{zero_nullity_expansion_formula}) where in $v \Phi_v$ certainly appears the term $\xi_a \Psi_a$ (and possibly other terms). By previous corollary applied to the term $\omega' = \myconjugate{v} \Phi_{\myconjugate{v}} + \omega''$ ($\omega''$ can be zero), it must contain $\xi_{-a} \Psi_{-a}$.
\end{proof}

\newpage

%

%
%



\end{document}